# Minimizing the effect of trends on detrended fluctuation analysis of long-range correlated noise

Radhakrishnan Nagarajan

Center on Aging, University of Arkansas for Medical Sciences

Rajesh G. Kavasseri

Department of Electrical and Computer Engineering, North Dakota State University

**Abstract:**

Detrended fluctuation analysis (DFA) has been proposed as a robust technique to determine possible long-range correlations in power-law processes [1]. However, recent studies have reported the susceptibility of DFA to trends [2] which give rise to spurious crossovers and prevent reliable estimation of the scaling exponents. Inspired by these reports, we propose a technique based on singular value-decomposition (SVD) of the trajectory matrix to minimize the effect of linear, power-law, periodic and also quasi-periodic trends superimposed on long-range correlated power-law noise. The effectiveness of the technique is demonstrated on publicly available data sets [2].

**PACS Code** : **05.40.-a**

**Keywords** : Detrended fluctuation analysis, Trends, Embedding, Singular Value Decomposition.

*Author for Correspondence*

Radhakrishnan Nagarajan
Center on Aging, University of Arkansas for Medical Sciences
629 Jack Stephens Drive, Room: 3105
Little Rock, Arkansas 72205
Phone: (501) 526 7461
Email: nagarajanradhakrish@uams.edu

# 1. Introduction

Power-law processes are ubiquitous and manifest themselves in a wide-range of complex systems. A partial list of examples include: physiological systems [3], biosequences [1], internet-traffic, [4] and stock-markets [5]. Power-law processes include the class of self-similar (self-affine) data sets that lack a well-defined temporal scale [3]. The nature of the correlation(s) in such processes can be studied by extracting its scaling exponent(s). While some processes (monofractals) can be quantified by a single scaling exponent ($\alpha$), others (multifractals) are characterized by a spectrum of exponents [6, 7]. Detrended fluctuation analysis (DFA) [1] and its extension multifractal DFA (MF-DFA) [8] are well known techniques to extract the scaling exponent(s). The ease of implementation and interpretation of results obtained from DFA/MF-DFA and its superiority over traditional estimators are some of the reasons behind its popularity. In the present study, we focus on long-range correlated monofractal data whose scaling exponent ($\alpha$) lies in the range $0.5 < \alpha \leq 1.0$ [9].

*Crossovers* [10] in the DFA log-log plot of the fluctuation function F(s) versus the time scale (s) have been reported to reflect change in correlation properties of the given data at different time scales. However, recent studies have reported that such crossovers can be a result of trends as opposed to varying scaling exponents of the power-law noise [2]. Such trends also prevent reliable estimation of the scaling exponent(s). Trends have been broadly classified to fall under linear, periodic and power-law forms [2]. Several factors can give rise to trends in data [2]. DC offsets and exponential decays in electric circuits are instances of linear and power-law trends. Periodic trends are attributed to the presence of *commensurate* frequencies in the data. An instance of periodic trends is seasonal variation. Quasi-periodic trends are accompanied by *incommensurate* frequencies and can occur as a result of perturbation of the periodic cycles [11].

This report is organized as follows. In Section 2, we discuss SVD based filtering of the linear, power-law, periodic and quasi-periodic trends. In Section 3, we show the effectiveness of the proposed techniques on publicly available data [2]. The data sets to be discussed include long-range correlated noise with scaling exponents ($\alpha = 0.8$) and ($\alpha = 0.9$) previously published in [2].

**2. SVD filters to minimize the effect of trends**

Singular value decomposition of a matrix $\Gamma_{pxq}$, $p>q$ is given by $\Gamma = U\Sigma V^T$, where $U_{pxq}$ and $V_{qxq}$ are the left and right orthogonal matrices and $\Sigma_{qxq}$, the diagonal matrix [12]. The diagonal elements of $\Sigma_{qxq}$ are the desired *singular values*, also known as *eigen-values*. The SVD of $\Gamma$ is related to the eigen-decomposition of the symmetric matrices $\Gamma^T\Gamma$ and $\Gamma\Gamma^T$, as $\Gamma^T\Gamma v_i = l_i^2 v_i$ and $\Gamma\Gamma^T u_i = l_i^2 u_i$. The non-zero eigen-values of $\Gamma^T\Gamma$ are the same as that of $\Gamma\Gamma^T$, and determine the rank of $\Gamma$. The singular values of $\Gamma$ are the square roots of the eigen-values of $\Gamma^T\Gamma$. The eigen-values are ordered such that $l_i > l_{i+1} \geq 0, \forall i$. The rank of $\Gamma$ is equal to the number of non-zero eigen-values. However, in the presence of numerical and measurement noise, if $l_i > 0, \forall i$ then, $\Gamma$ would be a full-rank matrix, with rank q. The close relation between SVD of $\Gamma$ and power spectral estimation is discussed below.

Power spectral techniques have proven to be powerful tools in identifying the dominant frequency components in the given data. Power spectral estimation can be broadly classified into parametric, non-parametric and subspace decomposition methods [13]. Subspace decomposition methods such as Pisarenko Harmonic Decomposition (PHD) [14], estimate the pseudospectrum of harmonically related sinusoids corrupted with zero mean additive white noise, by eigen-decomposition of the auto-correlation matrix. Techniques related to the PHD are Multiple SIgnal Classification (MUSIC) and Estimation of Signal Parameters via Rotational Invariance Technique (ESPRIT) [13]. The mathematical framework behind PHD relies on Caratheodory's uniqueness result (Appendix, A) [15, 16], which provides a bound on the size of the toeplitz matrix necessary to extract the $p$ distinct frequency components. Consider a toeplitz matrix of size $m$, with $m > p$, its eigen-decomposition yields $p$ dominant eigen-values and their corresponding eigen-vectors span the *signal subspace*. The *noise subspace* is spanned by the remaining ($m$-$p$) eigen-vectors. It might be interesting to note that the subspace decomposition can be achieved by the SVD of the trajectory matrix [17] with the embedding dimension ($d = m > p$) and time delay ($t = 1$). The normalized eigen-values obtained by the SVD of the trajectory matrix with embedding parameters ($d = 4$ and 20, $t = 1$) for periodic (Appendix, B3) and quasi-periodic data (Appendix, B4) is shown in Figure 1. The number of dominant frequency components reflected by the peaks (three) in the power spectrum correspond to $p = 6$. Therefore, a sharp decrease in the magnitude of the normalized eigen-values, $l_i^* = \log_e(\frac{l_i}{\sum l_i})$ is observed for ($d > 6$), Figure 1. This is true for the periodic (B3) as well as quasi-periodic (B4) data.

In the present study, we assume the deterministic trends superimposed on the power-law noise (Appendix, B) are reflected by dominant eigen-values in their eigen-value spectrum. The trends are also assumed to be uncorrelated to the power-law noise. The power-spectrum of a long-range correlated noise is broad-band and has a power-law form. While periodic and quasi-periodic trends are *stationary trends* and restricted to a small band, those of linear and power-law trends are *non-stationary* and can be *approximated* to a small band of frequencies in the broad band spectrum of the noise. In order to justify this assumption we investigated the eigen-value spectrum of the long-range correlated noise ($\alpha = 0.8$) to that of long-range correlated noise ($\alpha = 0.8$) superimposed with linear trend (Appendix B1), power-law trend with positive exponent (Appendix B21) and power-law trend with negative exponent (Appendix B22), Figure 2. The eigen-value spectrum revealed a considerable difference in the dominant eigen-value corresponding to ($p = 1$). Hence, for the subsequent filtering we chose ($p = 1$) for linear and power-law trends.

Traditional SVD filters retain the dominant *p* eigen-values and reject the (*m-p*) eigen-values on the noise floor. However, the objective here is to reject the components corresponding to the dominant eigen-values *p* and retain those (*m-p*) corresponding to the power-law noise. The dominant *p* eigen-values also have contributions from the noise. However, in the present study we assume that the contribution of the trend to the *p* dominant eigen-values is much greater than that of the noise. Thus, the p-dominant eigen-values in the SVD decomposition can be attributed largely due to the trend. As noted earlier, for periodic and quasi-periodic trends, it is necessary to reject the dominant *p* eigen-values corresponding to the dominant *p* frequency components. However, for linear and power-law trends, it is sufficient to reject the dominant eigen-value, i.e. $p = 1$.

As a comment, it should be noted that periodic and quasi-periodic trends may be a consequence of an interesting features superimposed on the power-law noise and can be generated by either linear or nonlinear processes. However, it is not possible to determine the nature of the process generating the trends from the given data [17]. In the following section, we present the SVD based algorithm to minimize the presence of various types of trends in long range correlated data.

*Algorithm*

**Given**: Power-law noise superimposed with linear, power-law, periodic or quasi-periodic trends $\{w_n\}$, $i = 1...N$ (Appendix, B).

**Step 1:** Embed $\{w_n\}$ with parameters ($d, t$) where $d$ is the embedding dimension and $t$ the time delay. The embedded data can be represented as a matrix $\Gamma$ given by:

$$\bm{g}_k = (w_k, w_{k+t}, ..., w_{k+N-(d-1)t}), 1 \leq k \leq d$$

$$\Gamma = \begin{bmatrix} \bm{g}_1 \\ . \\ . \\ . \\ \bm{g}_d \end{bmatrix} \quad (1)$$

The time delay is fixed at ($t = 1$), therefore $w_{ij} = w_{i+j-1}$, $1 \leq i \leq N - (d-1)t$ and $1 \leq j \leq d$.

For a power-law noise corrupted with:

- *Linear trend*: $p = 1$
- *Power-law trend*: $p = 1$

- *Periodic Trend*: $p$ = the number of dominant frequency components of the form $e^{jw_i t}, i = 1...p$ in their power-spectrum.
- *Quasi-periodic Trend*: $p$ = the number of dominant frequency components of the form $e^{jw_i t}, i = 1...p$ in their power-spectrum.

Choose the embedding dimension such that ($d \gg p$). Larger the choice of $d$, finer the representation of the power-law noise and better the filtering.

**Step 2**: Apply SVD to the matrix $\Gamma$, i.e. $\Gamma = U\Sigma V^T$. Let the number of frequency components in the periodic or quasi-periodic trend be $p$. Set the dominant $2p+1$ eigen-values in the matrix $\Sigma$ to zero, let the resulting matrix be $\Sigma^*$.

**Step 3**: The filtered matrix $\Gamma^*_{N-(d-1)t \ xd}$ determined as $\Gamma^* = U\Sigma^* V^T$ with elements $\{w_n^*\}$. This in turn is mapped back on to a one-dimensional or filtered data given by

$$w^*_{i+j-1} = w^*_{ij} \text{ where } 1 \leq i \leq N - (d-1)t \text{ and } 1 \leq j \leq d$$

## 3. Results

In the present study we use long-range correlated data sets with (N = 7168) samples with scaling exponents ($\alpha$ = 0.8) and ($\alpha$ = 0.9) [2]. While higher order polynomial detrending has been recently suggested [18, 19], we found that first-order DFA (DFA-1) with (N = 7168) samples was sufficient to extract the true scaling exponents ($\alpha$ = 0.8) and ($\alpha$ = 0.9). This was verified for the trend free data sets ($\alpha$ = 0.8, N = 7168) and ($\alpha$ = 0.9, N = 7168) before subsequent analysis. In order to justify that the choice of the sample size and first order regression is sufficient, we include the log-log plot of the fluctuation function (F(s)) versus the time-scale (s) as a reference in the following discussions. The parameters used for generating linear, power-law, periodic and quasi-periodic trends are included under Appendix, B.

In the following discussion, the notations introduced in (Appendix, B) will be used to denote long-range correlated power-law noise with $w_n$ and without trends $x_n^P$. The filtered data after applying the proposed algorithm is represented by $w_n^*$. For the power-law noise super-imposed with linear trend (Appendix B1) a considerable increase in the magnitude of the scaling exponent ($\alpha \sim 2$), Figure 3a, is observed failing to reflect the scaling exponent of the noise. Thus linear trends obscure the scaling exponent of the power-law noise ($\alpha = 0.8, 0.9$). The noise with trends $w_n$ is first embedded on to a high-dimensional space with parameters ($d = 10, 20, 40, 100$ with $t = 1$). The proposed algorithm is used to minimize the effect of trends with ($p = 1$). Increasing overlap in the log-log plot of the filtered data $w_n^*$ and the noise free data $w_n$ is observed with increasing embedding dimension, Figure 3. More importantly, with increasing embedding dimension, the fluctuation function at larger time-scales of the filtered data $w_n^*$, overlap with those of the noise free data $w_n$. However, spurious crossovers are still observed (Figures 3 a, b, c, d when $\alpha = 0.8$ and Figures 3 f, g, h, i when $\alpha = 0.9$) when smaller embedding dimensions are used. This behavior is consistent for all types of trends.

A similar behavior was observed for power-law trends with positive (Appendix, B2a-B2c, $p = 1$) and negative exponents superimposed on power-law noise, Figures 4-6. The scaling exponent was considerably greater in the presence on the data with power-law trends $w_n$, Figures 4(a, f) and Figures 5(a, f). With increasing embedding dimension ($d$) a considerable overlap was observed between $w_n^*$ and noise free data, Figures 4 and 5. In [2], it is shown that the scaling properties of correlated noise corrupted with trends can be determined from DFA computations on the noise and the trend, each considered separately, when the noise and trend are assumed to be uncorrelated. In the present context, the power-law trend (-0.7, N = 7168) would correspond to scaling exponent (1.5 - 0.7 = 0.8) which resembles that of the long-range correlated noise ($\alpha = 0.8$). In order to reject the claim that the observed filtering is an outcome of particular choice of the exponent of the power law trend, (-0.7), we examined the effectiveness of the filtering for power-law trend with exponent (-1.0, N = 7168). Following [2], a power-law exponent (-1.0) corresponds to scaling exponent of (1.5 - 1.0 = 0.5) which is different from the scaling exponent of the long-range correlated noise ($\alpha = 0.8$ and 0.9). The SVD filter effectively minimized the effect of the power-law trend (-1.0), Figure 6. Thus the performance of the filter is not affected by a particular choice of the scaling exponent.

For periodic (Appendix, B3, $p = 6$), Figure 1a, and quasi-periodic trends (Appendix, B4, $p = 6$), Figure 1c, apparent crossovers were observed in the data with trends $w_n$, Figures 7 (a, f) and 8 (a, f), consistent with earlier observations [2]. Considerable nonlinearity in the log-log plots of the fluctuation function versus time scale for these trends prevented the estimation of the scaling exponents. In all the types of trends considered here, it was consistently noted that increasing the embedding dimension facilitated reliable estimation of the true scaling exponent. The embedding dimension required for reliable estimation of the scaling exponent for the quasi-periodic data Figures 8 (d, i) was greater than that for the periodic trend, Figures 7 (d, i). This can be attributed to the fact that for periodic and harmonically related data, the power spectrum peaks at commensurate frequencies. The above results were confirmed with data sets with different parameters for linear, power-law, periodic and quasi-periodic trends. However, the results are not enclosed in the present report for brevity.

In order to determine the effect of reducing sample-size on the proposed filtering technique we chose data sets of length (N = 2048 with $\alpha$ = 0.8 and $\alpha$ = 0.9) which is about one-third of (N = 7168). While we were able to estimate the slopes ($\alpha$ = 0.8 and $\alpha$ = 0.9) from the log-log plot of F(s) vs s, for (N = 2048), reducing the sample-size introduced some nonlinearity in the log-log plots. Thus reducing the sample-size further might prevent reliable estimation of the scaling exponent even on the raw data. However, reduction in sample-size did not appreciably affect the performance of the SVD filter as discussed below. The reduced length data sets (N = 2048 with $\alpha$ = 0.8 and $\alpha$ = 0.9) were subsequently superimposed with linear, power-law, periodic and quasi-periodic trends with the same parameters used for (N = 7168), Section B. The effect of linear trends with varying embedding dimension is shown in Figure 9. A considerable part of the log-log plot of the filtered data overlapped with that of the raw data for ($d$ = 40). This has to be contrasted with the (N = 7168) which required ($d$ = 100) for a complete overlap. Thus reliable estimation in the presence of linear trends can be obtained for a smaller embedding dimension in the case of (N = 2048) as opposed to (N = 7168). A similar behavior was observed for the positive and negative power-law trends as shown in Figures 10 and 11. However, no appreciable change in the qualitative characteristics of the log-log plots was observed when (N = 7168) and (N = 2048) was super-imposed with periodic and quasi-periodic trends, Figures 12 and 13. Thus from the above discussion the proposed filtering procedure is robust to the variation in the sample-size. As a comment, it should be noted that the SVD procedure is a linear transform whose computational complexity increases with increasing sample-size which might restrict its application to all types of data.

**4. Discussion**

Detrended fluctuation analysis has been used to quantify long-range correlations in power-law noise across a wide range of monofractal and multifractal data. Recent reports have indicated the susceptibility of DFA to trends in the data. Several experimental factors can give rise to trends. Trends such as linear, power-law and periodic trends have been found to give rise to crossovers in the log-log plots of the fluctuation function versus time scale [2] and reflect spurious existence of more than a single scaling exponent at different time scales. These trends also prevent reliable estimation of the scaling exponent. Inspired by these reports, we suggest an SVD based filtering to minimize the effect of trends superimposed on long-range correlated noise by singular value decomposition of the corresponding trajectory matrix. The trends are assumed to be reflected by the dominant eigen-values in the corresponding eigen-value spectrum. The effectiveness of the SVD based filtering was demonstrated on linear, power-law, periodic and also quasi-periodic trends superimposed on long-range correlated noise. Long-range correlated noise is broad-band in nature whose power is distributed across the entire spectrum of frequencies. Periodic and quasi-periodic comprising of $p$ sinusoidal components are restricted to a small-band of frequencies and can be minimized by filtering the dominant $p$ components. Unlike periodic and quasi-periodic components, linear and power-law trends are non-stationary. While the power-spectrum is not defined for nonstationary trends, in the present study we observed that the effect of linear and power-law trends can be minimized by rejecting the dominant eigen-value ($p = 1$), (see Fig. 2). The present study implicitly assumes the trends to be deterministic and restricted to a narrow-band compared to the long-range correlated noise. However, trends can also be non-deterministic and can be encountered in real world data sets. In a recent study, the impact of signals with random spikes on DFA estimate was investigated [20]. Unlike the deterministic trends, non-deterministic trends are broad-band similar to that of the power-law noise and require further investigation. The effectiveness of the proposed technique is demonstrated on publicly available data sets. The log-log plot of the original data sets points with (DFA-1) reflected the true scaling exponents and was included as reference in all the discussion. We also observed that decreasing the sample-size did not appreciably affect the performance of the SVD filter.

## 5. Acknowledgements

We would like to thank the authors in Reference 2 for making available their data sets through Physionet. We would also like to thank the reviewers for their suggestions.

## Appendix

### A. Caratheodory's Uniqueness Result

Given a one-dimensional time series of the form $x_n = \sum_{k=1}^{p} |c_k|^2 e^{jw_k(n-1)}, n = 1...m$, if $m > p+1$, then $w_k \in [-p, p)$ and $|c_k|^2, k = 1...p$, are unique [15, 16].

### B. Trends

The long-range correlated power-law noise $x_n^P$ was superimposed with the periodic trends $x_n^T$ to yield the trended data $w_n = x_n^P + x_n^T, n = 1...7168$. In the present study, we assume that $x_n^P$ and $x_n^T$ are uncorrelated with one another.

B1. *Linear Trend*

$$x_n^T = 0.05n, \text{ where } n = 1...7168.$$

B2. *Power Law Trend*

B2a. $x_n^T = n^{0.4}$, where $n = 1...7168$.

B2b. $x_n^T = (\frac{n}{300})^{-0.7}$, where $n = 1...7168$.

B2c. $x_n^T = (\frac{n}{300})^{-1.0}$, where $n = 1...7168$.

B3. *Periodic Trend*

The sampling rate $F_s = 100$ was chosen so as to satisfy the Nyquist criterion.

$$f_1 = 2, f_2 = 15, f_3 = 31, F_s = 128$$

$$x_n^T = \sin(\frac{2pf_1 n}{F_s}) + \sin(\frac{2pf_2 n}{F_s}) - 2\sin(\frac{2pf_3 n}{F_s}) \text{ where } n = 1...7168.$$

B4. *Quasi-periodic trend*

The sampling rate $F_s = 100$ was chosen so as to satisfy the Nyquist criterion.

$$f_1 = \sqrt{2}, f_2 = \sqrt{15}, f_3 = \sqrt{31}, F_s = 128$$

$$x_n^T = \sin(\frac{2pf_1 n}{F_s}) + \sin(\frac{2pf_2 n}{F_s}) - 2\sin(\frac{2pf_3 n}{F_s}) \text{ where } n = 1...7168.$$

**Figures and Captions**

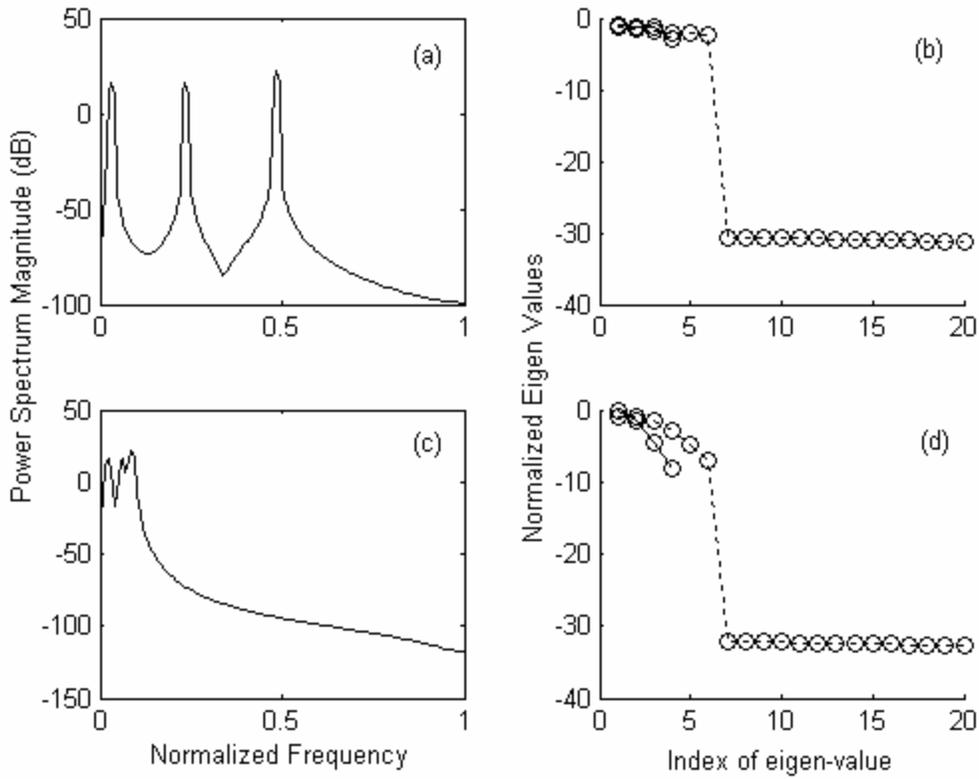

**Figure 1:** The power spectrum (a, c) and the corresponding normalized eigen-value spectrum, (b, d) obtained by SVD of the trajectory matrix for periodic (a, b, Appendix, B3) and quasi-periodic data (c, d, Appendix B4). Two different embedding dimensions ($d$ = 4 and 20, shown by the circles) with ($t$ = 1) were used to generate the trajectory matrix. A significant drop in the magnitude of the eigen-values is observed for ($d > 6$).

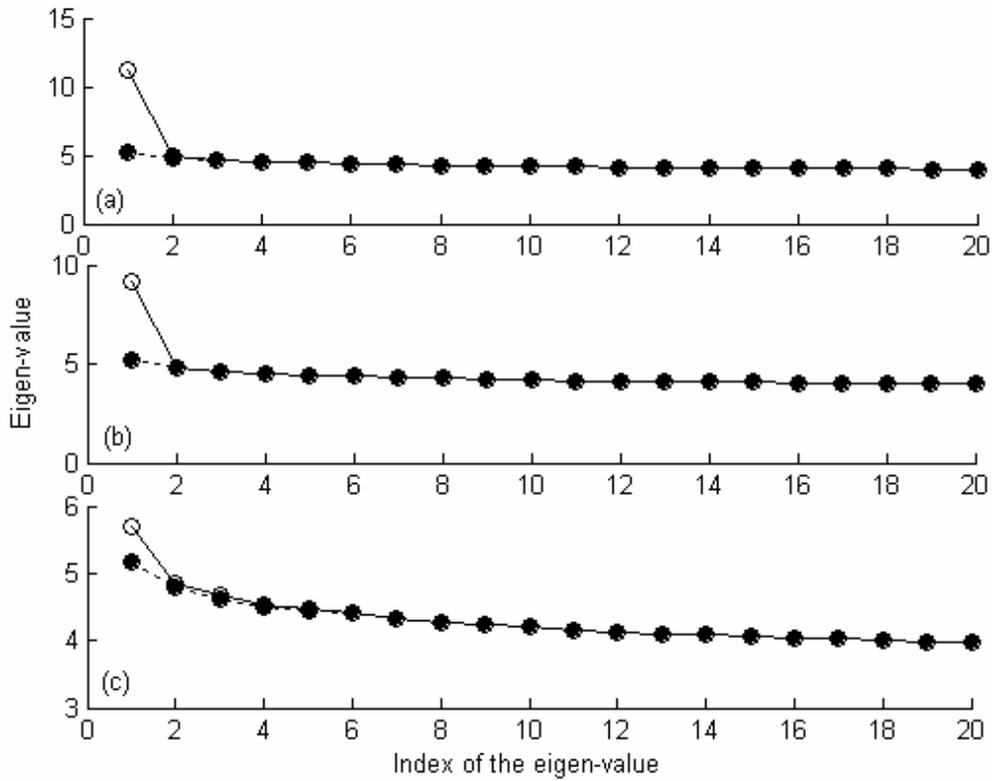

**Figure 2:** The eigen-value spectrum of the noise ($\alpha = 0.8$, solid circles) and noise superimposed with trends (open circles). Figures a, b and c correspond to linear trend (Appendix B1), power-law trend with positive exponent (Appendix B21) and power-law trend with negative exponent (Appendix B22) respectively. A significant difference in the magnitude of the dominant eigen-value is seen between the noise and noise superimposed with trends.

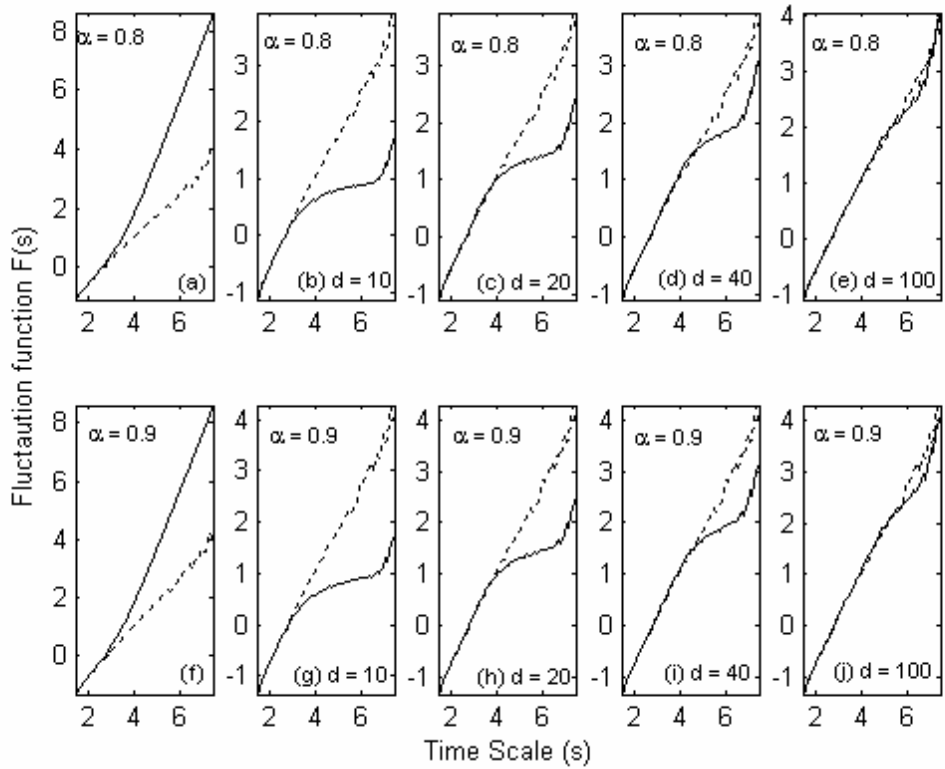

**Figure 3:** Log-log plots of the fluctuation function versus time scale ($\log_e F(s)$ vs $\log_e s$) obtained of the power-law noise (N= 7168) superimposed with linear trends (Appendix, B1) is shown in (a), $\alpha = 0.8$ and (f), $\alpha = 0.9$. The log-log plots, F(s) vs s, obtained by using the proposed algorithm with parameters ($d = 10, 20, 40, 100$, $\bm{t} = 1$, $p = 1$) for $\alpha = 0.8$ and $\alpha = 0.9$, is shown in plots (b, c, d, e) and (g, h, i, j) respectively. The log-log fluctuation plots of the original data ($\alpha = 0.8$ and $\alpha = 0.9$) without trends is shown (dashed lines) for reference inside each sub-plot.

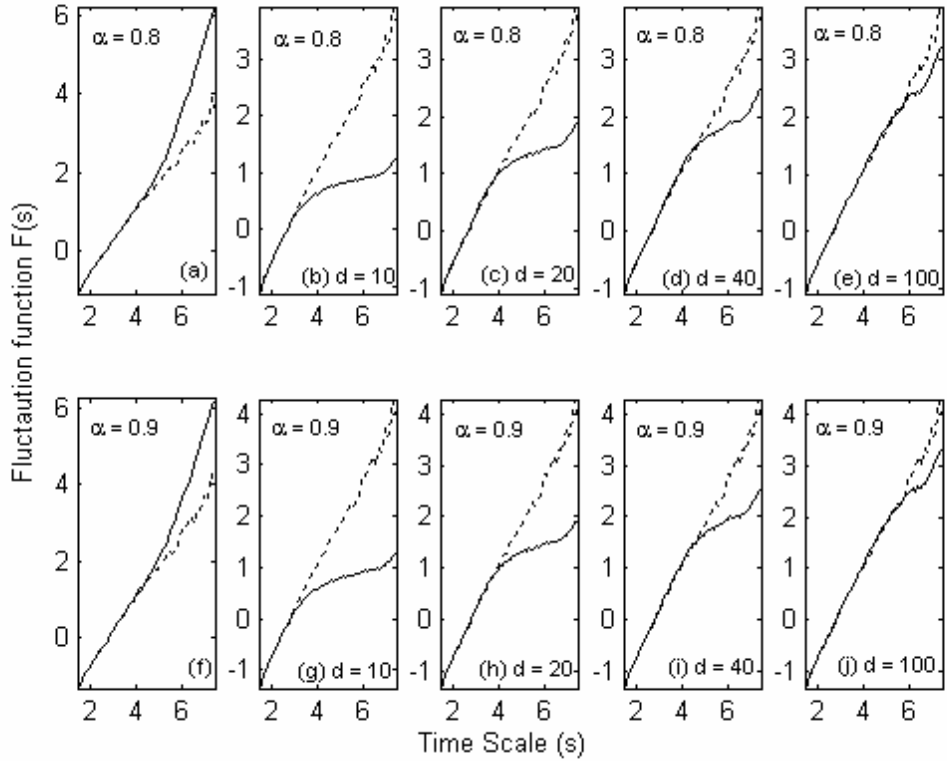

**Figure 4:** Log-log plots of the fluctuation function versus time scale ($\log_e F(s)$ vs $\log_e s$) obtained of the power-law noise (N= 7168) superimposed with power-law trends with positive exponent (Appendix, B2a) is shown in (a), $\alpha = 0.8$ and (f), $\alpha = 0.9$. The log-log plots, F(s) vs s, obtained by using the proposed algorithm with parameters ($d = 10, 20, 40, 100$, $t = 1$, $p = 1$) for $\alpha = 0.8$ and $\alpha = 0.9$, is shown in plots (b, c, d, e) and (g, h, i, j) respectively. The log-log fluctuation plots of the original data ($\alpha = 0.8$ and $\alpha = 0.9$) without trends is shown (dashed lines) for reference inside each sub-plot.

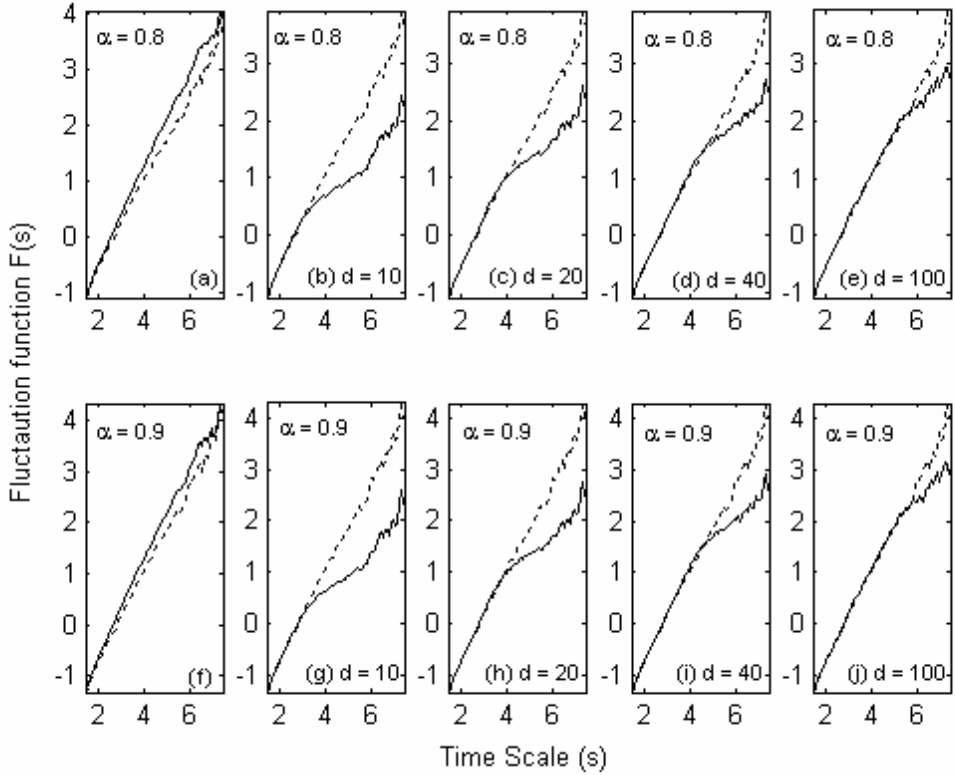

**Figure 5:** Log-log plots of the fluctuation function versus time scale ($\log_e F(s)$ vs $\log_e s$) obtained of the power-law noise (N = 7168) superimposed with power-law trends with negative exponent (Appendix, B2b) is shown in (a), $\alpha = 0.8$ and (f), $\alpha = 0.9$. The log-log plots, F(s) vs s, obtained by using the proposed algorithm with parameters ($d = 10, 20, 40, 100$ with $t = 1, p = 1$) for $\alpha = 0.8$ and $\alpha = 0.9$, is shown in plots (b, c, d, e) and (g, h, i, j) respectively. The log-log fluctuation plots of the original data ($\alpha = 0.8$ and $\alpha = 0.9$) without trends is shown (dashed lines) for reference inside each sub-plot.

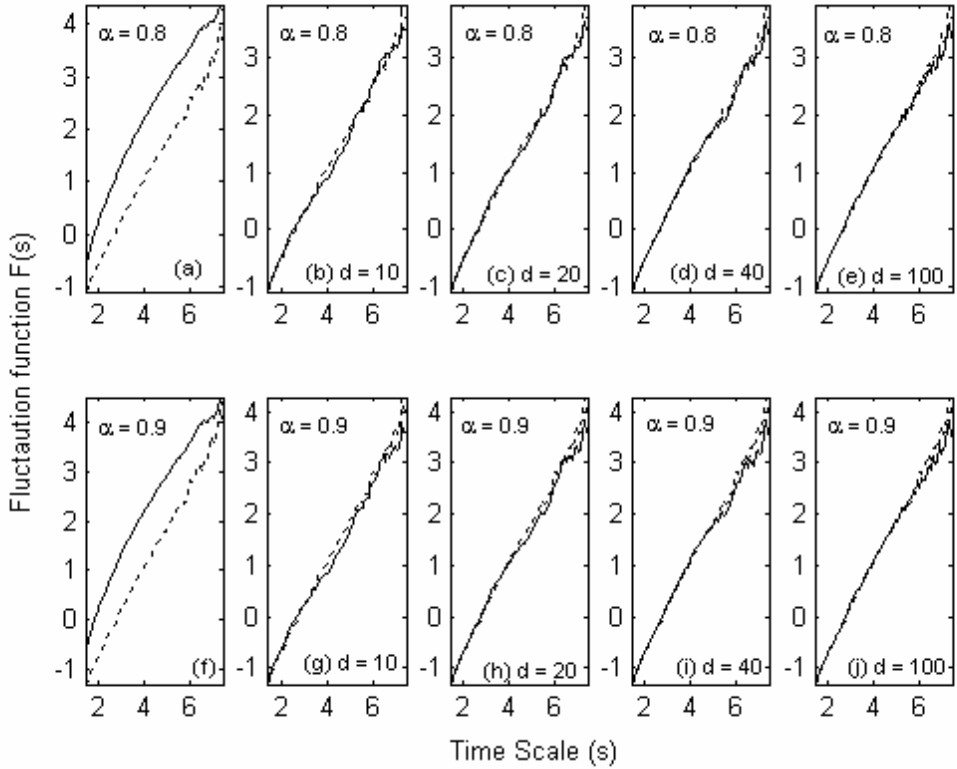

**Figure 6:** Log-log plots of the fluctuation function versus time scale ($\log_e F(s)$ vs $\log_e s$) obtained of the power-law noise (N = 7168) superimposed with power-law trends with negative exponent (Appendix, B2c) is shown in (a), $\alpha = 0.8$ and (f), $\alpha = 0.9$. The log-log plots, F(s) vs s, obtained by using the proposed algorithm with parameters ($d = 10, 20, 40, 100$ with $t = 1, p = 1$) for $\alpha = 0.8$ and $\alpha = 0.9$, is shown in plots (b, c, d, e) and (g, h, i, j) respectively. The log-log fluctuation plots of the original data ($\alpha = 0.8$ and $\alpha = 0.9$) without trends is shown (dashed lines) for reference inside each sub-plot.

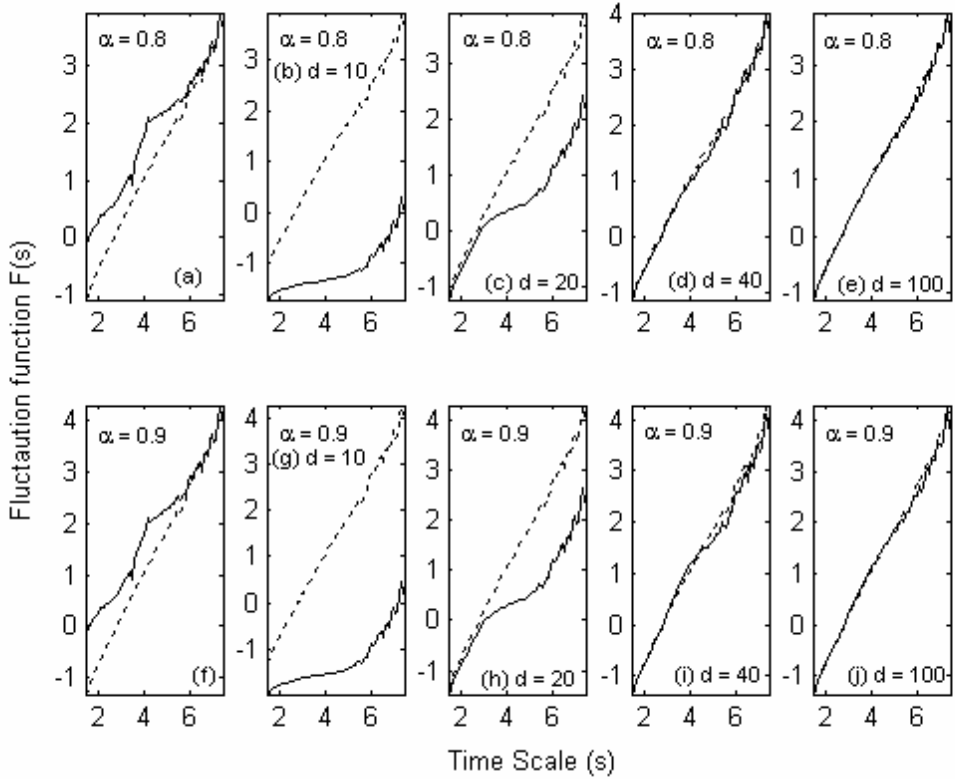

**Figure 7:** Log-log plots of the fluctuation function versus time scale ($\log_e F(s)$ vs $\log_e s$) obtained of the power-law noise (N = 7168) superimposed with periodic trends (Appendix, B3) is shown in (a), $\alpha = 0.8$ and (f), $\alpha = 0.9$. The log-log plots, F(s) vs s, obtained by using the proposed algorithm with parameters ($d = 10, 20, 40, 100$, $t = 1$, $p = 6$) for $\alpha = 0.8$ and $\alpha = 0.9$, is shown in plots (b, c, d, e) and (g, h, i, j) respectively. The log-log fluctuation plots of the original data ($\alpha = 0.8$ and $\alpha = 0.9$) without trends is shown (dashed lines) for reference inside each sub-plot.

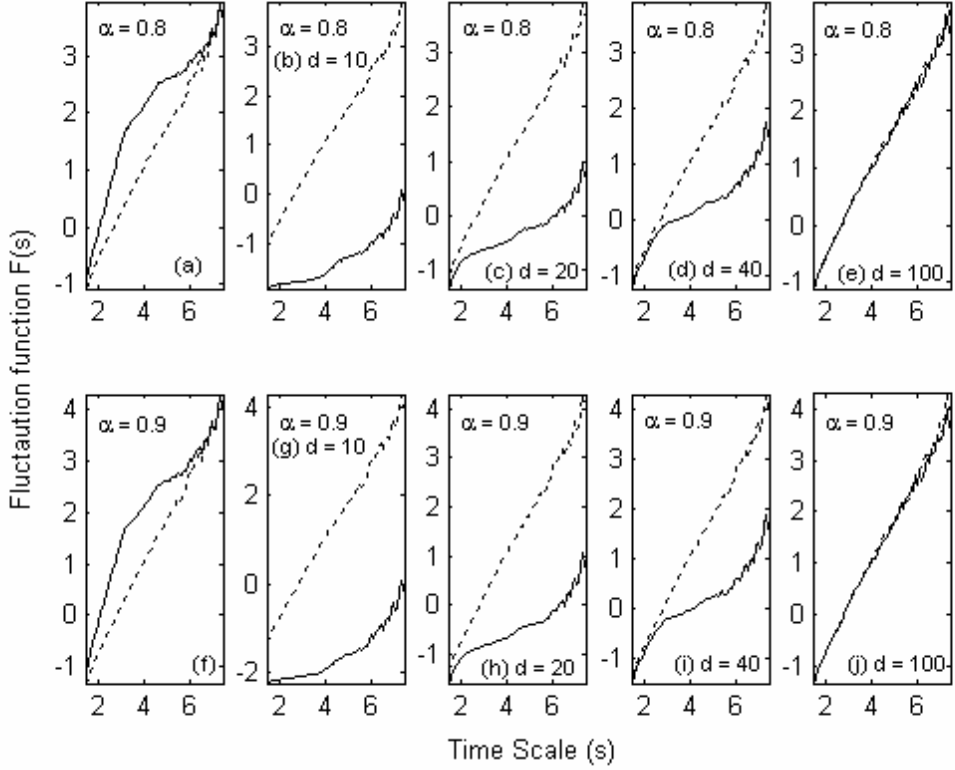

**Figure 8:** Log-log plots of the fluctuation function versus time scale ($\log_e F(s)$ vs $\log_e s$) obtained of the power-law noise (N = 7168) superimposed with quasi-periodic trends (Appendix, B4) is shown in (a), $\alpha = 0.8$ and (f), $\alpha = 0.9$. The log-log plots, F(s) vs s, obtained by using the proposed algorithm with parameters ($d$ = 10, 20, 40, 100, $t = 1$, $p = 6$) for $\alpha = 0.8$ and $\alpha = 0.9$, is shown in plots (b, c, d, e) and (g, h, i, j) respectively. The log-log fluctuation plots of the original data ($\alpha = 0.8$ and $\alpha = 0.9$) without trends is shown (dashed lines) for reference inside each sub-plot.

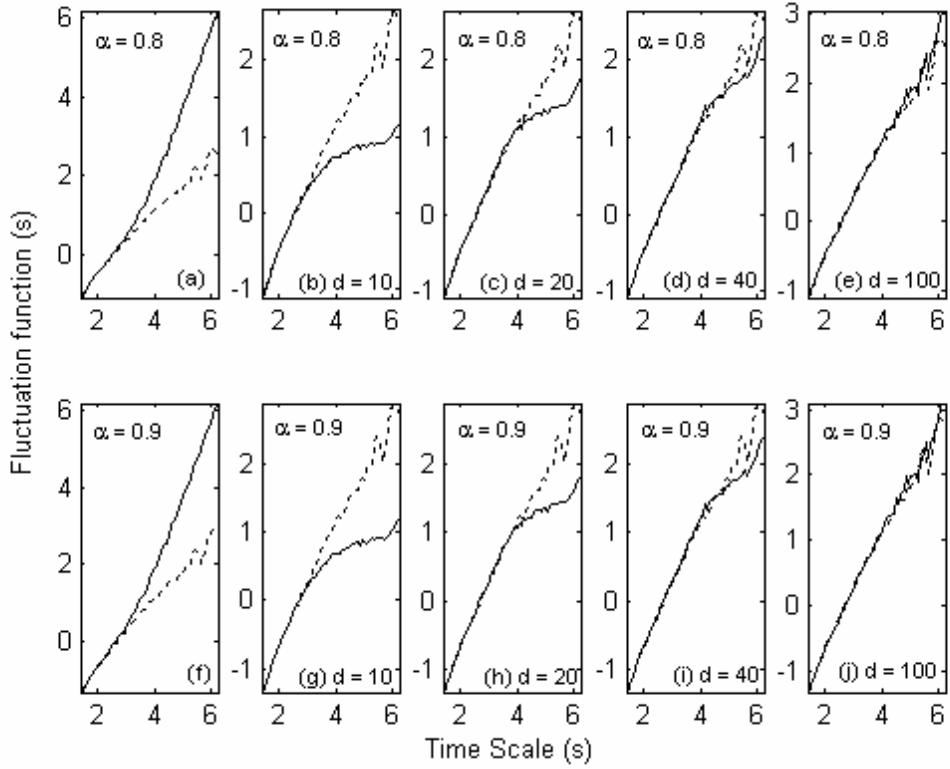

**Figure 9:** Log-log plots of the fluctuation function versus time scale ($\log_e F(s)$ vs $\log_e s$) obtained of the power-law noise (N= 2048) superimposed with linear trends (Appendix, B1) is shown in (a), $\alpha = 0.8$ and (f), $\alpha = 0.9$. The log-log plots, F(s) vs s, obtained by using the proposed algorithm with parameters ($d = 10, 20, 40, 100$, $t = 1$, $p = 1$) for $\alpha = 0.8$ and $\alpha = 0.9$, is shown in plots (b, c, d, e) and (g, h, i, j) respectively. The log-log fluctuation plots of the original data ($\alpha = 0.8$ and $\alpha = 0.9$) without trends is shown (dashed lines) for reference inside each sub-plot.

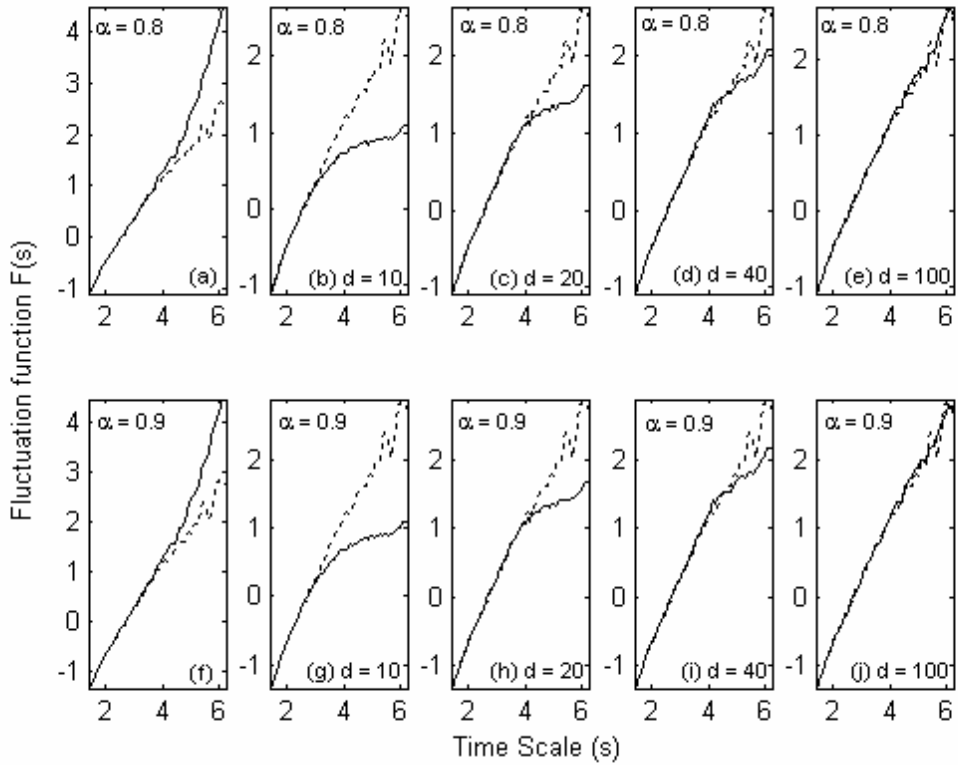

**Figure 10:** Log-log plots of the fluctuation function versus time scale ($\log_e F(s)$ vs $\log_e s$) obtained of the power-law noise (N= 2048) superimposed with power-law trends with positive exponents (Appendix, B2a) is shown in (a), $\alpha = 0.8$ and (f), $\alpha = 0.9$. The log-log plots, F(s) vs s, obtained by using the proposed algorithm with parameters ($d = 10, 20, 40, 100, t = 1, p = 1$) for $\alpha = 0.8$ and $\alpha = 0.9$, is shown in plots (b, c, d, e) and (g, h, i, j) respectively. The log-log fluctuation plots of the original data ($\alpha = 0.8$ and $\alpha = 0.9$) without trends is shown (dashed lines) for reference inside each sub-plot.

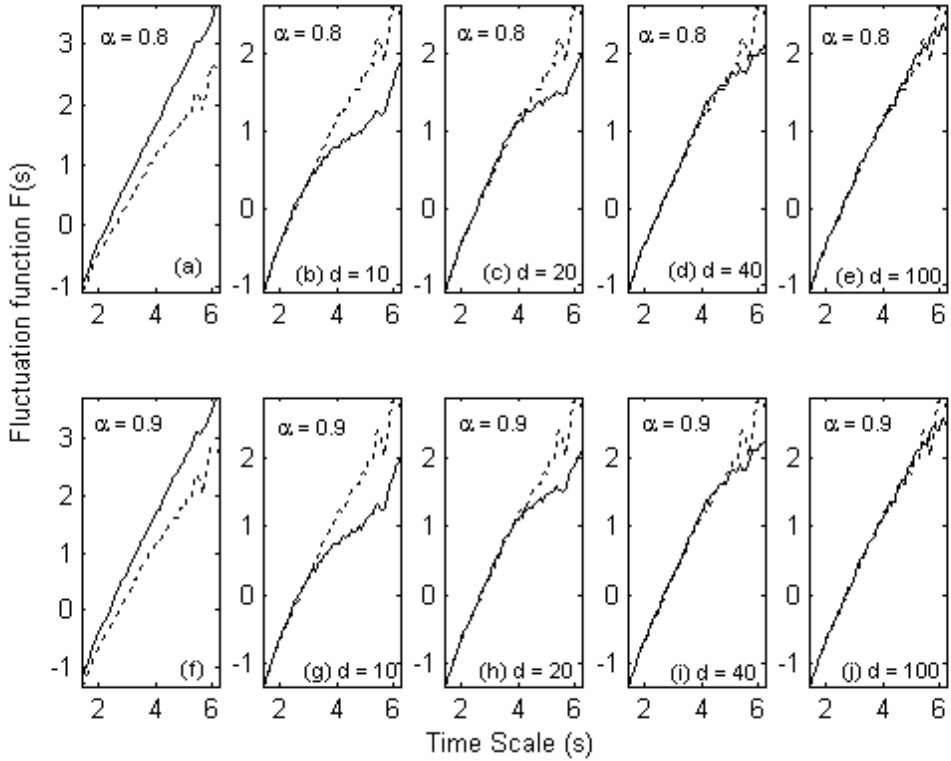

**Figure 11:** Log-log plots of the fluctuation function versus time scale ($\log_e F(s)$ vs $\log_e s$) obtained of the power-law noise (N = 2048) superimposed with power-law trends with negative exponents (Appendix, B2b) is shown in (a), $\alpha = 0.8$ and (f), $\alpha = 0.9$. The log-log plots, F(s) vs s, obtained by using the proposed algorithm with parameters ($d$ = 10, 20, 40, 100, $t$ = 1, $p$ = 1) for $\alpha = 0.8$ and $\alpha = 0.9$, is shown in plots (b, c, d, e) and (g, h, i, j) respectively. The log-log fluctuation plots of the original data ($\alpha = 0.8$ and $\alpha = 0.9$) without trends is shown (dashed lines) for reference inside each sub-plot.

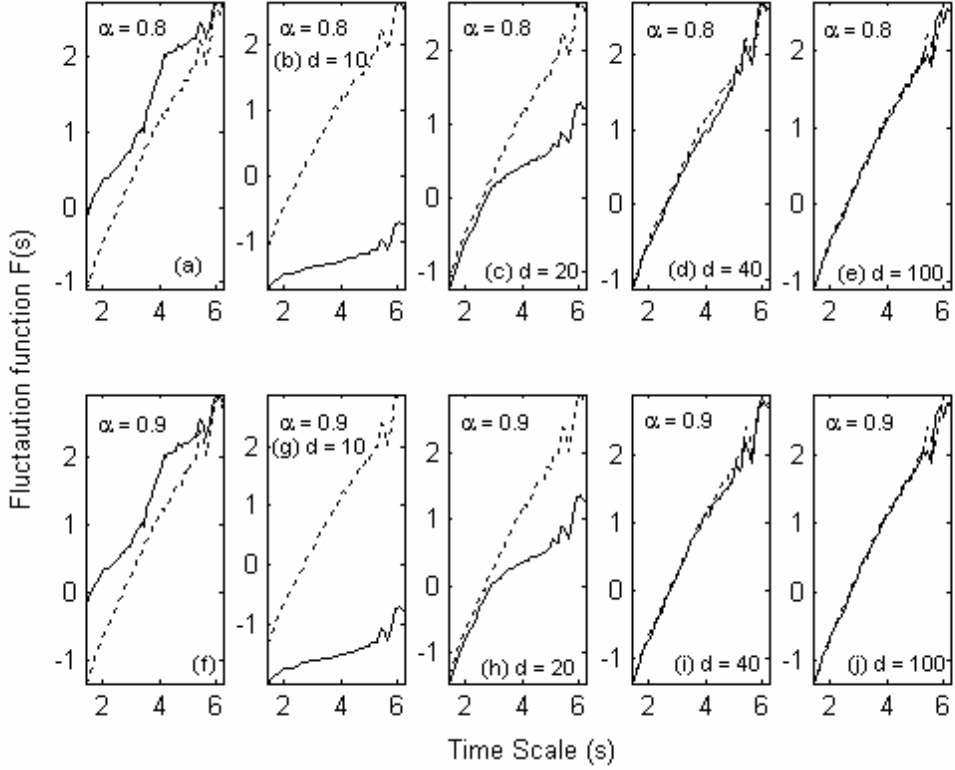

**Figure 12:** Log-log plots of the fluctuation function versus time scale ($\log_e F(s)$ vs $\log_e s$) obtained of the power-law noise (N = 2048) superimposed with periodic trends (Appendix, B3) is shown in (a), $\alpha = 0.8$ and (f), $\alpha = 0.9$. The log-log plots, F(s) vs s, obtained by using the proposed algorithm with parameters ($d$ = 10, 20, 40, 100, $t$ = 1, $p$ = 6) for $\alpha = 0.8$ and $\alpha = 0.9$, is shown in plots (b, c, d, e) and (g, h, i, j) respectively. The log-log fluctuation plots of the original data ($\alpha = 0.8$ and $\alpha = 0.9$) without trends is shown (dashed lines) for reference inside each sub-plot.

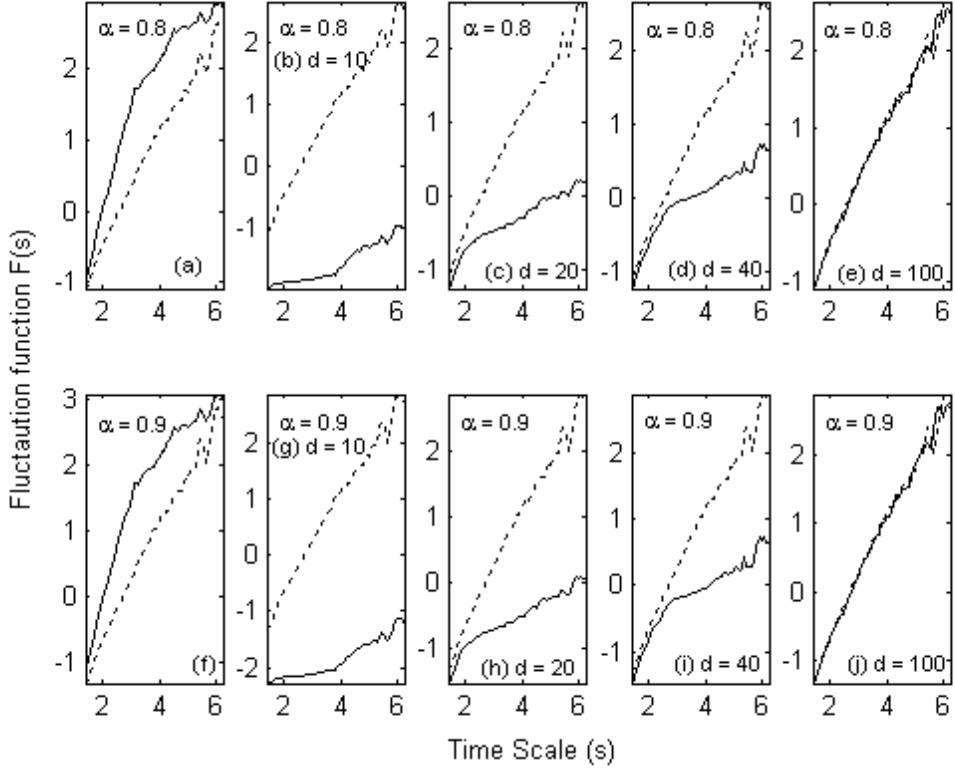

**Figure 13:** Log-log plots of the fluctuation function versus time scale ($\log_e F(s)$ vs $\log_e s$) obtained of the power-law noise (N = 2048) superimposed with quasi-periodic trends (Appendix, B4) is shown in (a), $\alpha = 0.8$ and (f), $\alpha = 0.9$. The log-log plots, F(s) vs s, obtained by using the proposed algorithm with parameters ($d = 10, 20, 40, 100$, $t = 1$, $p = 6$) for $\alpha = 0.8$ and $\alpha = 0.9$, is shown in plots (b, c, d, e) and (g, h, i, j) respectively. The log-log fluctuation plots of the original data ($\alpha = 0.8$ and $\alpha = 0.9$) without trends is shown (dashed lines) for reference inside each sub-plot.